# Cosmology with relativistically varying physical constants

Rajendra P. Gupta*

*Department of Physics, University of Ottawa, Ottawa, Canada  K1N 6N5*

September 15, 2020

**ABSTRACT**

We have shown that the varying physical constant model is consistent with the recently published variational approach wherein Einstein equations are modified to include the variation of the speed of light $c$, gravitational constant $G$ and cosmological constant $\Lambda$ using the Einstein-Hilbert action. The general constraint resulting from satisfying the local conservation laws and contracted Bianchi identities provides the freedom to choose the form of the variation of the constants as well as how their variations are related. When we choose $\dot{G}/G = 3\dot{c}/c$, $c = c_0 \exp[(a^\alpha - 1)]$, $G = G_0 \exp[3(a^\alpha - 1)]$ and $\Lambda = \Lambda_0 \exp[(a^{-\alpha} - 1)]$, where $a$ is the scale factor and $\alpha = 1.8$, we are able to show that the resulting model: (a) fits the supernovae 1a observational data marginally better than the $\Lambda$CDM model; (b) determines the first peak in the power spectrum of the cosmic microwave background temperature anisotropies at multipole value of $l = 217.3$; (c) calculates the age of the universe as 14.1 Gyr; and (d) finds the BAO acoustic scale to be 145.2 Mpc. These numbers are within less than 3% of the values derived using the $\Lambda$CDM model. Surprisingly we find that the dark-energy density is negative in a universe that has significant negative curvature and whose expansion is accelerating at a faster rate than predicted by the $\Lambda$CDM model.

**Key words:** cosmology: theory, cosmic background radiation, cosmological parameters, dark energy, dark matter, distance scale.

## 1. INTRODUCTION

Despite the stunning success of the standard model in explaining the cosmological observations, there remain many gaps that need to be bridged. We believe these gaps warrant looking at alternative theories based on examining the foundation of physics that has been developed over centuries from local observations extrapolated to the universe at large. One of the foundations is that constants relating various observables are indeed constants everywhere and at every time in the universe. This is especially so because attempts to measure their variation has put very tight limit on their potential variation.

Varying physical constant theories have been in existence since time immemorial (Thomson & Tate 1883; Weyl 1919; Eddington 1934) especially since Dirac (1937; 1938) suggested variation of the constant $G$ based on his large number hypothesis. Brans and Dicke (1961) developed the $G$ variation theory compliant with general relativity in which constant $G$ was raised to the status of scalar field potential. While Einstein developed his ground breaking theory of special relativity based on the constancy of the speed of light, he did consider its possible variation (Einstein 1907). This was followed by the varying speed of light theories by Dicke (1957), Petit (1988) Moffatt (1993a; 1993b). More recently, Albrecht and Magueijo (1999) and Barrow (1999) developed such a theory in which Lorentz invariance is broken as there is a preferred frame in which scalar field is minimally coupled to gravity. Other proposals include locally invariant theories (Avelino & Martins 1999; Avelino, Martins & Rocha 2000) and vector field theories that cause spontaneous violation of Lorentz invariance (Moffat 2016).

The most comprehensive review of the varying fundamental physical constants was done by Uzan (2003) followed by his more recent review (Uzan 2011). Chiba (2011) has provided an update of the observational and experimental status of the constancy of physical constants. We therefore will not attempt to cover the subject's current status except to mention a few of direct relevance to this work.

Variable physical constants are introduced in most of the proposed theories at the cost of either not conserving energy-momentum or violating Bianchi identity, which then leads to breaking the covariance of the theory. Such theories may be considered inconsistent or *ad hoc* (Ellis & Uzan 2005) or quasi-phenomenological (Gupta 2019). An action principle to take into account the variation of the fundamental constants that are being considered in a theory is required for generalization of Einstein equations. Costa et al. (Costa et al. 2019; Franzmann 2017) have attempted this approach by considering the speed of light $c$, gravitational constant $G$, and cosmological constant $\Lambda$ as scalar fields, and introducing an Einstein-Hilbert action that is considered consistent with the Einstein equations and with the general constraint that is compliant with contracted Bianchi identities and standard local conservation laws. Their approach is general covariant as it preserves the invariance of the general relativity. In the work of Costa et al. (2019) the focus was to explain early universe problems - the flatness problem and the horizon problem - without invoking inflation, which they have successfully shown by simply promoting the constants $G$ and $c$ in Einstein's equation to scalar fields. Here we will follow the same prescription. However, we will limit ourselves to consider important direct cosmological observations from now up to the time of cosmic microwave background emission.

We will begin with establishing the theoretical background for our work in this paper in Section 2 starting from the equations presented in the paper of Costa et al. (2019). We will confine ourselves to two considerations of the model: one with varying cosmological constant, and another with fixed cosmological constant. The standard formulae used in cosmology are based on fixed physical constants. We will go through carefully with the derivations of those we will use and determine how they are modified when the constants are allowed to evolve. These involve generalization of the luminosity distance, and therefore also of the distance modulus, as well as the scale factor for the surface of last scattering and the deceleration parameter.

Section 3 delineates the results. Firstly, we present the parameters and curves obtained by fitting the redshift versus distance modulus Pantheon Sample data (Scolnic et al. 2018) for

*Email: rgupta4@uottawa.ca

1048 supernovae 1a to the ΛCDM model and the two variable constants models. The second test is to see how well a model computes the multipole moment of the first peak in the power spectrum due to the temperature anisotropies in the cosmic microwave background. The third test may be considered how well the models match the value of the BAO acoustic scale obtained from the peak in the two point separation correlation function determined from measurements over a million galaxies. Another important test comprises determining the age of the universe.

Section 4 discusses the findings of this paper and Section 5 presents our conclusions.

## 2. THEORETICAL BACKGROUND

Following Costa et al. (2019) we may write the Einstein equations with varying physical constants (VPC) with respect to time $t$ - speed of light $c = c(t)$, gravitational constant $G = G(t)$ and cosmological constant $\Lambda = \Lambda(t)$ - applicable to the homogeneous and isotropic universe, as follows:

$$G_{\mu\nu} = \left(\frac{8\pi G(t)}{c(t)^4}\right) T_{\mu\nu} - \Lambda(t) g_{\mu\nu}. \quad (1)$$

Here $G_{\mu\nu} = R_{\mu\nu} - \frac{1}{2} g_{\mu\nu} R$ is the Einstein tensor with $R_{\mu\nu}$ the Ricci tensor and $R$ the Ricci scalar, and $T_{\mu\nu}$ is the stress energy tensor. Applying the contracted Bianchi identities, torsion free continuity and local conservation laws

$$\nabla^\mu G_{\mu\nu} = 0 \text{ and } \nabla^\mu T^{\mu\nu} = 0, \quad (2)$$

one gets a general constraint equation for the variation of the physical constants

$$\left[\frac{1}{G}\partial_\mu G - \frac{4}{c}\partial_\mu c\right]\left(\frac{8\pi G}{c^4}\right) T^{\mu\nu} - (\partial_\mu \Lambda) g^{\mu\nu} = 0. \quad (3)$$

Now the FLRW (Friedmann–Lemaître–Robertson–Walker) metric for the geometry of the universe is written as:

$$ds^2 = -c^2(t)dt^2 + a^2(t)\left[\frac{1}{1-kr^2}dr^2 + r^2(d\theta^2 + \sin^2\theta d\phi^2)\right], \quad (4)$$

with $k = -1, 0, 1$ depending on the spatial geometry of the universe.

The stress-energy tensor, assuming that the universe contents can be treated as perfect fluid, is written as:

$$T^{\mu\nu} = \frac{1}{c^2(t)}(\varepsilon + p)U^\mu U^\nu + pg^{\mu\nu}. \quad (5)$$

Here $\varepsilon$ is the energy density, $p$ is the pressure, $U^\mu$ is the 4-velocity vector with the constraint $g_{\mu\nu} U^\mu U^\nu = -c^2(t)$. (Unless necessary to avoid confusion, we will drop showing $t$ variation, e.g. $c(t)$ is written as $c$.)

Solving the Einstein equation (such as by using Maplesoft 2019) then yields VPC compliant Friedmann equations:

$$H^2 \equiv \frac{\dot{a}^2}{a^2} = \frac{8\pi G \varepsilon}{3c^2} + \frac{\Lambda c^2}{3} - \frac{kc^2}{a^2}, \Rightarrow \dot{a}^2 = a^2\left(\frac{8\pi G \varepsilon}{3c^2} + \frac{\Lambda c^2}{3} - \frac{kc^2}{a^2}\right), \quad (6)$$

$$\frac{\ddot{a}}{a} = -\frac{4\pi G}{3c^2}(\varepsilon + 3p) + \frac{\Lambda c^2}{3} + \frac{\dot{c}}{c}\frac{\dot{a}}{a} = -\frac{4\pi G}{3c^2}(\varepsilon + 3p) + \frac{\Lambda c^2}{3} + \frac{\dot{c}}{c}\frac{a}{\dot{a}}H^2. \quad (7)$$

Here a dot on top of a variable denotes the time derivative of that variable, e.g. $\dot{c} \equiv dc/dt$. Taking time derivative of Eq. (6), dividing by $2a\dot{a}$ and equating it with Eq. (7), yields the general continuity equation:

$$\dot{\varepsilon} + 3\frac{\dot{a}}{a}(\varepsilon + p) = -\left[\left(\frac{\dot{G}}{G} - 4\frac{\dot{c}}{c}\right)\varepsilon + \frac{c^4}{8\pi G}\dot{\Lambda}\right] \quad (8)$$

Eq. (3) for the FLRW metric and perfect fluid stress-energy tensor reduces to:

$$\left[\left(\frac{\dot{G}}{G} - 4\frac{\dot{c}}{c}\right)\frac{8\pi G}{c^4}\varepsilon + \dot{\Lambda}\right] = 0, \quad (9)$$

therefore,

$$\dot{\varepsilon} + 3\frac{\dot{a}}{a}(\varepsilon + p) = 0. \quad (10)$$

Using the equation of state relation $p = w\varepsilon$ with $w = 0$ for matter and $w = 1/3$ for relativistic particles, the solution for this equation is $\varepsilon = \varepsilon_0 a^{-3-3w}$, where $\varepsilon_0$ is the current energy density of all the components of the universe when $a = a_0 = 1$.

Next we need to consider the continuity equation Eq. (9). The simplest solution is by assuming $\Lambda$ =constant. Then $\frac{\dot{G}}{G} = 4\frac{\dot{c}}{c}$, i.e. $\frac{G}{G_0} = \frac{c^4}{c_0^4}$. This is what Costa et al. (2019) used in their paper. We will label this model as VcG model (varying $c$ and $G$ model). In fact they dropped the cosmological constant altogether for their $c$-flation solution. However, one could choose any relationship between $G$ and $c$, say $\frac{\dot{G}}{G} = \sigma \frac{\dot{c}}{c}$. Then from Eq. (9), by defining $\varepsilon_\Lambda = \frac{c^4 \Lambda}{8\pi G}$, we have

$$\frac{8\pi G}{c^4}\frac{\dot{c}}{c}(4-\sigma)\varepsilon = \dot{\Lambda}, \Rightarrow \frac{\dot{c}}{c} = \frac{c^4 \Lambda}{8\pi G}\left(\frac{\dot{\Lambda}}{\Lambda}\right)\left(\frac{1}{(4-\sigma)\varepsilon}\right) \equiv \frac{\varepsilon_\Lambda}{(4-\sigma)\varepsilon}\frac{\dot{\Lambda}}{\Lambda},$$

$$\Rightarrow \varepsilon_\Lambda = \frac{\dot{c}}{c}\frac{\Lambda}{\dot{\Lambda}}(4-\sigma)\varepsilon \quad (11)$$

The parameter $\sigma$ may be determined based on the physics or by fitting the observations. We have determined in the past (Gupta 2019) that $\sigma = 3$, i.e. $\frac{\dot{G}}{G} = 3\frac{\dot{c}}{c}$ and confirmed it by fitting the SNe 1a data. Thus, we must have $\varepsilon_\Lambda = \frac{\dot{c}}{c}\frac{\Lambda}{\dot{\Lambda}}\varepsilon$. We will label this model as VcGΛ model (varying $c$, $G$ and $\Lambda$ model).

The most common way of defining the variation of the constant is by using the scale factor powerlaw (Barrow & Magueijo 1999; Salzano & Dabrowski 2017) such as $c = c_0 a^\alpha$ which results in $\frac{\dot{c}}{c} = \alpha \frac{\dot{a}}{a} = \alpha H$. The advantage is that it results in very simple Freedman equations. However, as $a \to 0$ the variable constant tends to zero or infinity depending on the sign of $\alpha$. So, it yields reasonable results when $a = \frac{1}{1+z}$ corresponds to relatively small redshift $z$, but not for large $z$. What we have used in the past (Gupta 2019) is the relations like $\frac{\dot{c}}{c} = \alpha H_0$, i.e. $c = c_0 \exp[\alpha H_0(t - t_0)]$, which leads to a limiting value of $c$ at $t = 0$. This approach was very useful when solving problems with $t - t_0$ very small, such as for explaining astrometric anomalies. However, due to the involvement of time coordinate, it is difficult to use it in very simple Friedmann equations for their general solution. We therefore tried here another relation that results in:

$$c = c_0 \exp[(a^\alpha - 1)]; G = G_0 \exp[3(a^\alpha - 1)]; \text{ and }$$
$$\Lambda = \Lambda_0 \exp[(a^\beta - 1)]. \quad (12)$$

Their limitation is that $c$ can decrease in the past at most by a factor of $e = 2.7183$ and $G$ can decrease by a factor of $e^{-3}$ (for positive $\alpha$ within the region of their applicability). In the limit of $a \to 1$, they reduce to the earlier exponential forms. For example $\dot{c} = c_0 \exp(a^\alpha - 1) \times \alpha a^{\alpha-1} \dot{a}$, or $\frac{\dot{c}}{c} = \alpha a^\alpha \frac{\dot{a}}{a}$, which in the limit of $a \to 1$ is $\frac{\dot{c}}{c} = \alpha H_0$, or $c = c_0 \exp[\alpha H_0 (t - t_0)]$. It should be mentioned that we found in an earlier work (Gupta 2018) that analytically $\alpha = 1.8$.

Using relations of Eq. (12), we can now write Eq. (11) for $\sigma = 3$,

$$\frac{c^4 \Lambda}{8\pi G} \equiv \varepsilon_\Lambda = \frac{\dot{c}}{c} \frac{\Lambda}{\dot\Lambda} \varepsilon = \frac{\alpha}{\beta} a^{\alpha-\beta} \varepsilon. \tag{13}$$

The first Friedmann equation, Eq. (6), becomes

$$H^2 = \frac{8\pi G}{3c^2}\left(\varepsilon + \frac{\Lambda c^4}{8\pi G}\right) - \frac{kc^2}{a^2} = \frac{8\pi G}{3c^2}(\varepsilon + \varepsilon_\Lambda) - \frac{kc^2}{a^2}$$
$$= \frac{8\pi G}{3c^2}\varepsilon\left(1 + \frac{\alpha}{\beta}a^{\alpha-\beta}\right) - \frac{kc^2}{a^2}. \tag{14}$$

Dividing by $H_0^2$, and since $\varepsilon = \varepsilon_m + \varepsilon_r = \varepsilon_{m,0} a^{-3} + \varepsilon_{r,0} a^{-4}$ where subscript $m$ is for matter and $r$ is for radiation (relativistic particles, e. g. photons and neutrinos), we get

$$\frac{H^2}{H_0^2} = \frac{8\pi G}{3c^2 H_0^2}(\varepsilon_{m,0}a^{-3} + \varepsilon_{r,0}a^{-4})\left(1 + \frac{\alpha}{\beta}a^{\alpha-\beta}\right) - \frac{kc^2}{H_0^2 a^2}. \tag{15}$$

At $t = t_0$ (current time), $a = 1$ and $H = H_0$. Therefore,

$$1 = \frac{8\pi G_0}{3c_0^2 H_0^2}(\varepsilon_{m,0} + \varepsilon_{r,0})\left(1 + \frac{\alpha}{\beta}\right) - \frac{kc_0^2}{H_0^2}$$
$$= (\Omega_{m,0} + \Omega_{r,0})\left(1 + \frac{\alpha}{\beta}\right) - \frac{kc_0^2}{H_0^2}. \tag{16}$$

Here we have defined the current critical density as $\varepsilon_{c,0} = 3c_0^2 H_0^2/8\pi G_0$, $\Omega_{m,0} = \varepsilon_{m,0}/\varepsilon_{c,0}$ and $\Omega_{r,0} = \varepsilon_{r,0}/\varepsilon_{c,0}$. Thus, by defining $\Omega_0 = (\Omega_{m,0} + \Omega_{r,0})\left(1 + \frac{\alpha}{\beta}\right)$, we may write Eq. (16)

$$\Omega_{k,0} \equiv -\frac{kc_0^2}{H_0^2} = 1 - \Omega_0, \tag{17}$$

$$\frac{H^2}{H_0^2} = \exp[(a^\alpha - 1)](\Omega_{m,0}a^{-3} + \Omega_{r,0}a^{-4})\left(1 + \frac{\alpha}{\beta}a^{\alpha-\beta}\right) + \Omega_{k,0}\exp[2(a^\alpha - 1)]a^{-2} \equiv E(a)^2. \tag{18}$$

Now, the FLRW metric, Eq. (4), can be written in its alternative form (Ryden 2017) as

$$ds^2 = -c^2 dt^2 + a(t)^2[dr^2 + S_k(r)^2(d\theta^2 + \sin^2\theta d\phi^2)]. \tag{19}$$

Here $S_k(r) = R\sin(r/R)$ for $k = +1$ (closed universe); $S_k(r) = r$ for $k = 0$ (flat universe), $S_k(r) = R\sinh(r/R)$ for $k = -1$ (open universe), where $R$ is the parameter related to the curvature. The proper distance $d_P$ between an observer and a source is determined at fixed time by following a spatial geodesic at constant $\theta$ and $\phi$. Then

$$ds = a(t)dr \Rightarrow d_P(t) = a(t)\int_0^r dr = a(t)r. \tag{20}$$

We could determine $r$ following a null geodesic from the time $t$ a photon is emitted by the source to the time $t_0$ it is detected by the observer with $ds = 0$ in Eq. (19) at constant $\theta$ and $\phi$:

$$c^2 dt^2 = a(t)^2 dr^2 \Rightarrow \frac{cdt}{a(t)} = dr \Rightarrow r = \int_0^r dr = \int_t^{t_0}\frac{cdt}{a(t)}$$
$$\Rightarrow d_P = a(t_0)\int_t^{t_0}\frac{cdt}{a(t)}. \tag{21}$$

We may write $dt = da \cdot dt/da = da/\dot{a} = da/a\dot{a}/a = da/aH$, and $a = 1/(1+z)$, $da = -dz/(1+z)^2 = -a^2 dz$. Therefore

$$dt = -\frac{adz}{H} = -\frac{adz}{\frac{H_0 H}{H_0}} = -\frac{adz}{H_0 E(a)}, \text{ and} \tag{22}$$

$$d_P = \frac{1}{H_0}\int_0^z \frac{cdz}{E(z)} = \frac{c_0}{H_0}\int_0^z \frac{\exp[((1+z)^{-\alpha} - 1)]dz}{E(z)}. \tag{23}$$

Here $E(z)$ is obtained from Eq. (18) by substituting $1/(1+z)$ for $a$.

*Constant $\Lambda$ Model:* Let us also consider the proposition of Costa et al. (2019) that $\Lambda$ be taken as constant (albeit for the study of very early universe) - the VcG model. While they did not propose how $c$ and $G$ should vary in general, we will use the same form as for the VcG$\Lambda$ model so that the results of the two can be comparable. In that case, as mentioned above, we have

$$c = c_0 \exp[(a^\alpha - 1)], G = G_0 \exp[4(a^\alpha - 1)], \text{ and } \Lambda = \Lambda_0. \tag{24}$$

Then, the first Friedmann equation becomes:

$$H^2 = \exp[2(a^\alpha - 1)]\left[\frac{8\pi G_0}{3c_0^2}(\varepsilon_{m,0}a^{-3} + \varepsilon_{r,0}a^{-4}) + \frac{\Lambda_0 c_0^2}{3} - kc_0^2 a^{-2}\right]. \tag{25}$$

Then, as before, we get

$$\frac{H^2}{H_0^2} = \exp[2(a^\alpha - 1)]\left[\Omega_{m,0}a^{-3} + \Omega_{r,0}a^{-4} + \Omega_{\Lambda,0} + \Omega_{k,0}a^{-2}\right] \equiv E(a)^2. \tag{26}$$

Substituting $a = 1/(1+z)$ we get $E(z)^2$. We can use Eq. (23) with this $E(z)$ to obtain proper distance corresponding to the VcG model.

*$\Lambda$CDM Model:* If we substitute $\alpha = 0$ in Eq. (26), we get back the case of physical constants not varying. We can then solve for the standard $\Lambda$CDM model parameters by fitting the observational data, or put the parameter of our choice to see how well they fit the data.

### 2.1 Redshift vs. Distance Modulus

We will now consider the specifics of applying the theory to develop relationship between the redshift $z$ and distance modulus $\mu$. First of all we will check if the relation $a = 1/(1+z)$ holds when physical constants are varying.

Since light travels along null geodesics, we have as per Eq. (21) $c^2 dt^2 = a(t)^2 dr^2$, or $dr = cdt/a(t)$. A wave crest of the light emitted from a galaxy at time $t_e$ and observed at time $t_0$ travels a distance

$$r = \int_0^r dr = \int_{t_e}^{t_0}\frac{cdt}{a(t)}. \tag{27}$$

Next wave crest will be emitted at time $t_e + \lambda_e/c_e$ and observed at time $t_0 + \lambda_0/c_0$ and will cover the same distance

$$r = \int_0^r dr = \int_{t_e + \lambda_e/c_e}^{t_0 + \lambda_0/c_0}\frac{cdt}{a(t)}. \tag{28}$$

Therefore,

$$\int_{t_e}^{t_0} \frac{cdt}{a(t)} = \int_{t_e+\lambda_e/c_e}^{t_0+\lambda_0/c_0} \frac{cdt}{a(t)}. \quad (29)$$

Subtracting from both sides the integral $\int_{t_e+\lambda_e/c}^{t_0} \frac{cdt}{a(t)}$ we get

$$\int_{t_e}^{t_e+\lambda_e/c_e} \frac{cdt}{a(t)} = \int_{t_0}^{t_0+\lambda_0/c_0} \frac{cdt}{a(t)}. \quad (30)$$

Assuming $c$ and $a(t)$ remain unchanged over the extremely short time between two consecutive wave crests compared to the age of the universe, we have

$$\frac{c_e}{a_e} \int_{t_e}^{t_e+\lambda_e/c_e} dt = \frac{c_0}{a_0} \int_{t_0}^{t_0+\lambda_0/c_0} dt \Rightarrow \frac{c_e}{a_e}\left(\frac{\lambda_e}{c_e}\right) = \frac{c_0}{a_0}\left(\frac{\lambda_0}{c_0}\right) \Rightarrow \frac{\lambda_e}{a_e} = \frac{\lambda_0}{a_0}. \quad (31)$$

Since $z \equiv (\lambda_0 - \lambda_e)/\lambda_e$, we get from above $1 + z = a_0/a_e = 1/a_e$ with $a_0 \equiv 1$. This result is the same as for the case when $c$ is not varying.

Now the distance of a source is determined by measuring flux of radiation arriving from the source with known luminosity. We therefore need to relate proper distance $d_P$ that relates to the redshift $z$ as per Eq. (23) to the luminosity distance $d_L$ which relates to the flux. We will then be able to establish what we mean by luminosity distance in the variable physical constant approach and how it differs from the case when constants are indeed constants.

The photons emitted by a source at time $t_e$ are spread over the sphere of proper radius $S_k(d_P)$ and, referring to Eq. (19), proper surface area $A_P(t_0)$ is given by (Ryden 2017)

$$A_P(t_0) = 4\pi S_k(d_P)^2. \quad (32)$$

The flux is defined as luminosity $L$ divided by the area in a stationary universe. When the universe is expanding then the flux is reduced by a factor $1 + z$ due to energy reduction of the photons. We need to also determine how the increase in distance between the emitted photons affects the flux.

The proper distance between two emitted photons separated by a time interval $\delta t_e$ is $c_e \delta t_e$ whereas the proper distance between the same two photons when observed would become $(c_0 \delta t_e)(1 + z)$. Thus the proper distance would increase by a factor $(c_0 \delta t_e)(1 + z)/c_e \delta t_e = (1 + z)c_0/c_e$. When $c$ is constant, i.e. $c_e = c_0$, this effect reduces the flux by a factor of $1 + z$. But when it is not, there is an additional factor $c_0/c_e$ that we have to consider to correct the flux. We thus have to account for an extra factor $c_0/c_e$ in the luminosity distance increase when calculating proper distance of a source from its flux data.

The flux $f$ and the luminosity distance $d_L$ relation may therefore be written as

$$f = \frac{L}{4\pi S_k(d_P)^2 (1+z)^2 \left(\frac{c_0}{c_e}\right)}, \quad (33)$$

$$d_L \equiv \left(\frac{L}{4\pi f}\right)^{1/2} = S_k(d_P)(1+z)\left(\frac{c_0}{c_e}\right)^{1/2}. \quad (34)$$

The distance modulus $\mu$ by definition is related to the luminosity distance $d_L$:

$$\mu \equiv 5\log_{10}\left(\frac{d_L}{1Mpc}\right) + 25$$

$$= 5\log_{10} S_k(d_P) + 5\log_{10}(1+z) + 25 + 2.5\log_{10}\left(\frac{c_0}{c_e}\right) \quad (35)$$

The last term is the extra correction term that must be included in calculating the distance modulus when one is considering the varying speed of light irrespective of the model applied and whether or not the space is flat.

### 2.2 CMB Power Spectrum

Here we will be focussing on the first peak in the power spectrum of the cosmic microwave background (CMB) temperature anisotropies and the related parameters such as its multipole moment, sound horizon distance at the time universe became transparent, acoustic scale derived from baryonic acoustic oscillation, and angular diameter distance of the surface of last scattering.

The sound horizon distance $d_s(t_{ls})$ is the distance sound travels at speed $c_s(t)$ in photon-baryon fluid from the big-bang until the time such plasma disappeared due to the formation of the atoms, i.e. the time of last scattering $t_{ls}$. Following Eq. (21) we may write (Durrer 2008)

$$d_{sh}(t_{ls}) = a(t_{ls}) \int_0^{t_{ls}} \frac{c_s(t)dt}{a(t)}. \quad (36)$$

The speed of sound $c_s(t)$ in terms of the speed of light $c(t)$ in the photon-baryon fluid with baryon density $\Omega_b$ and radiation density $\Omega_r$ is given by (Durrer 2008)

$$c_s(t) \approx \frac{c(t)}{\sqrt{3}}\left(1 + \frac{3\Omega_b}{4\Omega_r}\right)^{-1/2}. \quad (37)$$

Substituting it in Eq. (36) and making use of Eq. (23), we get

$$d_{sh}(z_{ls}) = \frac{c_0}{\sqrt{3}H_0(1+z_{ls})} \int_\infty^{z_{ls}} \frac{\exp[((1+z)^{-\alpha}-1)]dz}{\left(1+\frac{3\Omega_b}{4\Omega_r}\right)^{1/2} E(z)}. \quad (38)$$

This distance represents the maximum distance over which the baryon oscillations imprint maximum fluctuations in thermal radiation that is observed as anisotropies in CMB. This represents an angular size $\theta_{sh}$ observed at an angular diameter distance $d_A(z_{ls})$ given simply by

$$D_A(z_{ls}) \equiv S_k(d_P)/(1+z_{ls}), \text{ and } \theta_{sh} \equiv d_{sh}/D_A \quad (39)$$

And the corresponding multipole moment is given by $l = \pi/\theta_{sh}$.

Next thing is to determine the correct value of $z_{ls}$ corresponding to the last scattering surface when the physical constants are varying. The blackbody radiation energy density $\varepsilon_\gamma$ is given by (Ryden 2017) $\varepsilon_\gamma = 8\pi^2 k_B^4 T^4/15h^3 c^3$ ($k_B$ being the Boltzmann constant), and the radiation energy density as per Eq. (10) is given by $\varepsilon_r = \varepsilon_0 a^{-4}$. Therefore, when the constants $k_B, h$ and $c$ are not varying then $T \propto a^{-1}$ with $a^{-1} = 1 + z$. Using our earlier finding that the $h$ varies as $c$ (Gupta 2019), and that $k_B$ varies as $c^{1.25}$ as discussed in Section 4 of this paper, we see that $T \propto a^{-1}\exp[0.25(a^\alpha - 1)]$. Additionally, we have to see how does the ionization energy $Q$ of an atom evolve since it is proportional to $R_\infty hc$, where $R_\infty \equiv m_e e^4/8\epsilon_0^2 h^3 c$ with $m_e$ the mass of electron, $e$ its charge and $\epsilon_0 \propto 1/c^2$ the permittivity of space. Thus $R_\infty$ does not vary and therefore $Q \propto hc$. And then how does the thermal photon energy $kT$ evolve, since the ratio of the ionization energy and thermal energy determines the temperature of the last scattering surface? This ratio can be seen to evolves as $(hc/k_B) \propto \exp[0.75(a^\alpha - 1)]$. Cumulatively, the two effect lead to $T \propto$ $a^{-1}\exp[(a^\alpha - 1)] \equiv a'^{-1}$. Now $\exp[(1+z)^{-\alpha} - 1] = 1/e$ for $z \gg 1$. Therefore $T = T_0 a'^{-1} = T_0 (ae)^{-1}$. Thus $a'^{-1} = 1 + z' = (1+z)/2.7183$. We can now see that when $z = 1089, z' = 400$, i.e. the last scattering surface is at the redshift 400 in our VPC models.

One may question that the blackbody spectrum will be affected due to the varying physical constants since the energy density for the photons in the frequency range $\nu$ and $\nu + d\nu$ is given by $\varepsilon(\nu)d\nu = \frac{8\pi h\nu^3}{c^3}[d\nu/(\exp(\frac{h\nu}{k_B T}) - 1)]$, which would evolve due to the variation of $c, h$ and $k_B$. However, we do not know what exactly the spectrum of the distant cosmological objects is. We only know what it is when it is observed. If we consider the peak photon energy of the spectrum, it is given by (Ryden 2017) $h\nu_{max} = 2.8k_B T$, or $\nu_{max} = \frac{2.8k_B}{h}T = \frac{2.8k_{B,0}}{h_0}T_0 \exp[1.25(a^{1.8} - 1)]/a$. Then $\dot{\nu}_{max} = \frac{2.8k_0}{h_0}T_0\exp(1.25(a^{1.8} - 1))\left(\frac{1}{a}\right)(2.25a^{1.8} - 1)\left(\frac{\dot{a}}{a}\right)$. Therefore $\frac{\dot{\nu}_{max}}{\nu_{max}} = (2.25a^{1.8} - 1)\dot{a}/a$, which, near $a = a_0 = 1$, i.e. $t = t_0$ is $\frac{\dot{\nu}_{max}}{\nu_{max}} = 1.25H_0$. Since $H_0 \approx 71$ km s$^{-1}$Mpc$^{-1} = 2.3 \times 10^{-18}s^{-1}$, we get $\frac{\dot{\nu}_{max}}{\nu_{max}} \approx 3 \times 10^{-18}s^{-1} \approx 9 \times 10^{-11}$yr$^{-1}$. This is very small shift to observe in the blackbody spectrum. Nevertheless, it would be nice if one could come up with some observation or experiment which could detect the thermal spectrum variation with time.

### 2.3 Other Cosmological Parameters

Let us now consider some other cosmological parameters.

*Deceleration Parameter:* The second Friedmann equation, Eq. (7) is used to determine the deceleration parameter $q_0$, which by definition (Narlikar 2002)

$$q_0 = -\left(\frac{\ddot{a}}{aH^2}\right)_{t=t_0} = -\frac{1}{H_0^2}\left(\frac{\ddot{a}}{a}\right)_{t=t_0}. \tag{40}$$

Recalling that $a_0 = 1$ and $\dot{c}/c = \alpha H_0$ in the limit of $t = t_0$, we may write Eq. (7) at $t = t_0$ as

$$\left(\frac{\ddot{a}}{a}\right)_{t=t_0} = -\frac{4\pi G_0}{3c_0^2}\varepsilon_0(1 + 3w) + \frac{\Lambda_0 c_0^2}{3} + H_0\left(\frac{\dot{c}}{c}\right)_{t=t_0}, \text{ or}$$

$$q_0 = -\frac{\ddot{a}_0}{H_0^2} = \frac{8\pi G_0}{3c_0^2 H_0^2}\left[\varepsilon_0\left(\frac{1}{2} + \frac{3}{2}w\right) - \frac{\Lambda_0 c_0^4}{8\pi G_0}\right] - \alpha. \tag{41}$$

Since $\varepsilon_{c,0} = \frac{8\pi G_0}{3c_0^2 H_0^2}$, we get

$$q_0 = \frac{1}{2}\Omega_{m,0} + \Omega_{r,0} - \Omega_{\Lambda,0} - \alpha. \tag{42}$$

The last term results from the varying physical constants approach.

*Age of the Universe:* We need to calculate the parameter $t_0$ that is the current cosmic time relative to the big-bang time. From Eq. (18), for VcG$\Lambda$ model

$$\frac{H}{H_0} = E(a) \Rightarrow \frac{\dot{a}}{a} = H_0 E(a) \Rightarrow \frac{da}{dt} = H_0 aE(a)$$
$$\Rightarrow H_0 dt = \frac{da}{aE(a)} \equiv \frac{da}{F(a)}, \tag{43}$$

$$F(a)^2 = \exp[(a^\alpha - 1)](\Omega_{m,0}a^{-1} + \Omega_{r,0}a^{-2})\left(1 + \frac{\alpha}{\beta}a^{\alpha-\beta}\right) + \Omega_{k,0}\exp[2(a^\alpha - 1)], \tag{44}$$

$$\therefore t_0 = \int_0^{t_0} dt = \frac{1}{H_0}\int_0^1 da/F(a). \tag{45}$$

*BAO Acoustic Scale:* It relates to baryon acoustic oscillations (BAO) that are linked directly to the CMB anisotropies. BAO is considered observable today through the correlation function of galaxies' distribution in space (Anderson et al. 1914). It is given by:

$$r_{as} = d_{sh}(z_{ls})(1 + z_{ls}). \tag{46}$$

Here $d_{sh}$ is given by Eq. (35) and $z_{ls} = 400$ for the VPC models. This parameter could be considered another test for the VPC models.

## 3 RESULTS

We will now test the proposed model against observations. The most used test is to see how a model fits the redshift - distance modulus $(z - \mu)$ data from the observation on supernovae 1a (standard candle). The data used in this work is the so-called Pantheon Sample of 1048 supernovae Ia in the range of $0.01 < z < 2.3$ (Scolnic et al.). The data is in terms of the apparent magnitude and we added 19.35 to it to obtain normal distance modulus numbers as suggested by Scolnic (private communication).

The Matlab curve fitting tool was used to fit the data by minimizing $\chi^2$ and the latter was used for determining the corresponding $\chi^2$ probability $P$ (Press et al. 1992). Here $\chi^2$ is the weighted summed square of residual of $\mu$

$$\chi^2 = \sum_{i=1}^N w_i[\mu(z_i; H_0, p_1, p_2 \dots) - \mu_{obs,i}]^2, \tag{47}$$

where $N$ is the number of data points, $w_i$ is the weight of the $i$th data point $\mu_{obs,i}$ determined from the measurement error $\sigma_{\mu_{Obs,i}}$ in the observed distance modulus $\mu_{obs,i}$ using the relation $w_i = 1/\sigma_{\mu_{Obs,i}}^2$, and $\mu(z_i; H_0, p_1, p_2..)$ is the model calculated distance modulus dependent on parameters $H_0$ and all other model dependent parameter $p_1, p_2$, etc. As an example, for the $\Lambda$CDM models considered here, $p_1 \equiv \Omega_{m,0}$ and there is no other unknown parameter.

We then quantified the goodness-of-fit of a model by calculating the $\chi^2$ probability for a model whose $\chi^2$ has been determined by fitting the observed data with known measurement error as above. This probability $P$ for a $\chi^2$ distribution with $n$ degrees of freedom (DOF), the latter being the number of data points less the number of fitted parameters, is given by:

$$P(\chi^2, n) = \left(\frac{1}{\Gamma(\frac{n}{2})}\right)\int_{\chi^2/2}^\infty e^{-u}u^{\frac{n}{2}-1}du, \tag{48}$$

where $\Gamma$ is the well know gamma function that is generalization of the factorial function to complex and non-integer numbers. The lower the value of $\chi^2$, the better the fit, but the real test of the goodness-of-fit is the $\chi^2$ probability $P$; the higher the value of $P$ for a model, the better the model's fit to the data. We used an online calculator to determine $P$ from the input of $\chi^2$ and DOF (Walker 2020).

Our primary findings are presented in Figure 1 and Table 1.

One additional thing we wished to explore was if the prior of $\sigma = 3$ is the right choice for the VcG$\Lambda$ model. This can be confirmed by fitting the SNe Ia data. Thus, with all the parameters $(H_0, \Omega_{m,0}, \alpha, \beta, \sigma)$ free, i.e. no priors, we obtained results closed to the values in Table 1 but with large 95% confidence bounds. When we progressively constrained the parameters based on our prior knowledge, the 95% confidence bounds shrank and we obtained those parameters still unconstrained as expected. The results are presented in Table 2.

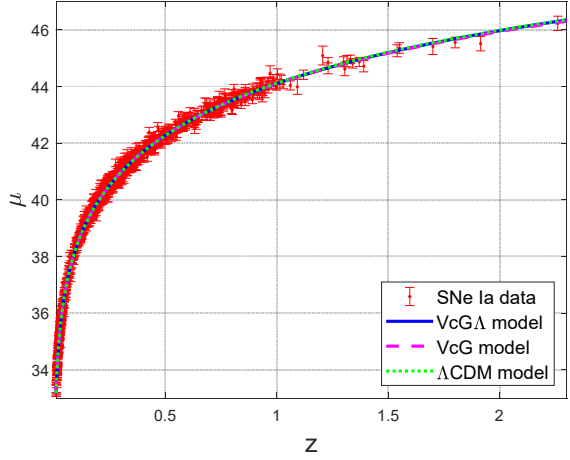

**Figure 1.** Supernovae Ia redshift $z$ vs. distance modulus $\mu$ data fits using the VcGΛ model and VcG model as compared to the fit using the ΛCDM model. All the curves are so close to each other that they appear to be superimposed.

**Table 1**. Parameters for the three models determined by fitting the Pantheon Sample SNe Ia $z - \mu$ data. This table shows all the models are able to fit the data very well with the VcGΛ model showing slight edge on others. The unit of $H_0$ is km s$^{-1}$ Mpc$^{-1}$. P% is the $\chi^2$ probability in percent; the higher the $\chi^2$ probability $P$, the better the model fits to the data. $R^2$ is the square of the correlation between the response values and the predicted response values. RMSE is the root mean square error and DOF is the degrees of freedom.

| Parameter | ΛCDM | VcG | VcGΛ |
|---|---|---|---|
| $H_0$ | 70.18±0.42 | 70.79±0.70 | 70.87±0.50 |
| $\Omega_{m,0}$ | 0.2845±0.0245 | 1.053±0.1363 | 0.528±0.037 |
| $\Omega_{\Lambda,0}$ | 1-$\Omega_{m,0}$ | 0.06441±0.22419 | -$\Omega_{m,0}$ |
| $\Omega_0$ | 1 | 1.117 | 0 |
| $\alpha$ | NA | 1.25 | 1.8 |
| $\beta$ | NA | NA | -1.8 |
| $\chi^2$ | 1036 | 1032 | 1032 |
| DOF | 1046 | 1045 | 1046 |
| P% | 58.1 | 60.7 | 61.5 |
| R-sq | 0.997 | 0.997 | 0.997 |
| RMSE | 0.9951 | 0.9938 | 0.9933 |

We notice that as the parameters are progressively fixed, the $\chi^2$ value does not change. This means that the constraint put on the parameter is reasonable, otherwise $\chi^2$ value would have increased. In fact, the $\chi^2$ probability $P$ slightly improves and so does the RMSE. SNe Ia data fit therefore confirms our choice of $\sigma = 3$ and $\alpha = 1.8$ for the VcGΛ model based on our previous work.

The SNe Ia test is to see how good a model is for the directly observable universe for which the redshift and light fluxes of galaxies are measurable. This test is the primary test. If the model fails this test then it may not be worthwhile to consider it for other tests.

**Table 2.** This table shows the results of fitting the Pantheon Sample SNe Ia $z - \mu$ data in order to determine the value of the parameter $\sigma$ that is used in the VcGΛ model. The description of other parameters is the same as in Table 1. The unit of $H_0$ is km s$^{-1}$ Mpc$^{-1}$.

| Parameter | All free parameters | 4 free parameters | 3 free parameters | 2 free parameters |
|---|---|---|---|---|
| $H_0$ | $70.47^{-1059}_{+1200}$ | 70.87 Fixed | 70.87 Fixed | 70.87 Fixed |
| $\Omega_{m,0}$ | $0.5273^{-20.86}_{+21.91}$ | $0.5211^{0.3566}_{0.6857}$ | $0.5298^{0.3537}_{0.7058}$ | $0.528^{0.4989}_{0.5575}$ |
| $\alpha$ | $1.85^{-35.94}_{+39.64}$ | $1.854^{0.769}_{2.938}$ | 1.8 Fixed | 1.8 Fixed |
| $\beta$ | $-2.29^{-125.9}_{+121.4}$ | $-2.204^{-14.8}_{+10.39}$ | $-1.756^{-5.701}_{+2.189}$ | $-1.8$ Fixed |
| $\sigma$ | $3.027^{-71.17}_{+77.22}$ | $3.001^{2.942}_{3.059}$ | $3.001^{2.956}_{3.045}$ | $3^{2.973}_{3.028}$ |
| $\chi^2$ | 1032 | 1032 | 1032 | 1032 |
| DOF | 1043 | 1044 | 1045 | 1046 |
| P% | 58.98 | 59.83 | 60.67 | 61.51 |
| R-sq | 0.997 | 0.997 | 0.997 | 0.997 |
| RMSE | 0.9948 | 0.9943 | 0.9938 | 0.9933 |

The second test is to see how well the model computes the multipole moment of the first peak in the power spectrum due to the temperature anisotropies in the cosmic microwave background. WMAP (Wilkinson Microwave Anisotropy Probe) and Plank (Planck Collaboration 2019) have determined the multipole moment $l \approx 220$ for the first peek.

A third test may be considered how well the models match the value of the BAO acoustic scale and compare it with the measured (Anderson et al. 2014) and Planck estimate of this parameter. To be consistent with the measured value using Hubble constant (Anderson et al. 2014) of $H_0 = 70$ km s$^{-1}$ Mpc$^{-1}$, we have normalized the two other models to $H_0 = 70$ km s$^{-1}$ Mpc$^{-1}$ for computing BAO acoustic scale only.

Another important test comprises determining the age of the universe. We have determined it for the models and compared it with the well accepted value of 13.7 Gyr.

Table 3 presents the above tests. It also includes deceleration parameter $q_0$, as well as some other cosmological parameters for ready comparison. Both the VPC models estimate higher acceleration than the ΛCDM model.

## 4. DISCUSSION

An objective of this paper is to see if the theory of Costa et al. (2019) complements our quasi-phenomenological model and provides improvement on it by being covariantly relativistic. The resulting very simple Friedmann equations and continuity equation eliminates the arbitrariness of our earlier models (Gupta 2018; Gupta 2019). The continuity equation, Eq. (8) essentially breaks down into three separate equations: the first for energy density $\varepsilon$, the second relating speed of light $c$ to the gravitational constant $G$, and the third relating the energy density $\varepsilon$ to the dark-energy density through the cosmological constant Λ. As a result the solution of the Freedmann equations is greatly simplified even though the physical constants $c, G$ and Λ are allowed to vary. Despite its simplicity, the VcGΛ model is highly satisfactory in explaining the observations from $z = 0$ to $z = 1100$. It is therefore worth examining it for other cosmological attributes.

**Table 3.** This table presents the results of other tests for the two models and compare them to the ΛCDM fit derived from the CMB temperature anisotropy observations by Planck (Planck Collaboration 2019). The Hubble constant used for the other two models is the same as obtained for these models from SNe Ia data fit. VSL stands for varying speed of light. The unit of $H_0$ is km s$^{-1}$ Mpc$^{-1}$.

| Parameter | | Planck+BAO ΛCDM | VcG | VcGΛ |
|---|---|---|---|---|
| Matter energy density | $\Omega_{m,0}$ | 0.3106 | 1.053 | 0.528 |
| Dark-energy density | $\Omega_{\Lambda,0}$ | 0.6894 | 0.06441 | -0.528 |
| Curvature parameter | $\Omega_{k,0}$ | -0.0096 | -0.1175 | 1.0000 |
| Hubble constant | $H_0$ | 67.70 | 70.79 | 70.87 |
| Redshift at last scattering surface | z | 1090 | 1090 | 1090 |
| VSL corrected redshift (see text) | z' | 1090 | 400 | 400 |
| 1st peak's multipole moment | $l$ | 220.6 | 130 | 217.3 |
| Deceleration parameter | $q_0$ | -0.5341 | -0.788 | -1.01 |
| Age of the universe | $t_0$ (Gyr) | 13.787 | 6.068 | 14.108 |
| BAO acoustic scale (measured value 149.28) | $r_s$ (Mpc) | 147.57 | 180.6 | 145.2 |

The first test both the VcG and VcGΛ models passed is the SNe 1a test (which involves relatively low redshift data) as is clear from studying Table 1 and Figure 1. However, as is obvious from Table 3, the VcG model did not pass other tests including the age of the universe test and those which involve high redshift observations whereas VcGΛ model passed them all with flying colors. Few noticeable things about the VcGΛ model in the table are: (a) the sign of the dark-energy density is negative, (b) the curvature parameter $\Omega_{k,0} = 1$ meaning that the space is curved negatively unlike in the ΛCDM model that yields the space as flat, and (c) the deceleration parameter is almost twice as much as for the ΛCDM model meaning that the universe's expansion is accelerating at almost twice the rate predicted by the ΛCDM model. These discrepancies are a result of the varying physical constants.

The staggering feature of the varying physical constant model VcGΛ is that it yields negative dark-energy density while still showing accelerated expansion of the universe. Possibility of such a scenarios has been explored by many researchers (e.g. Visinelli, Vagnozzi & Danielsson 2019; Dutta et al. 2020; Hartle, Hawking & Hertog 2012; Wang et al. 2018) who considered the varying cosmological constant to primarily resolve Hubble constant tension, which is due to the difference in its values obtained from CMB data and SNe Ia data. Such a scenario is extremely interesting from the string theory perspective as obtaining a vacuum solution with a positive value of Λ within moduli-fixed consistent and stable string theory compactifications has been a formidable task (Maldacena & Nunez 2001; Kachru et al. 2003; Conlon & Quevedo 2007; Danielson & van Riet 2018).

It is important to point out that the negative cosmological constant does not yield satisfactory outcome in most models. For example, Visinelli et al (2019) have considered one such model to see if their negative cosmological constant model, that is consistent with string theory, could resolve Hubble constant tension between its value determined from SNe Ia data and CMB anisotropy spectrum. Their model comparison analysis determined that the ΛCDM model is favoured over their negative cosmological constant model. However, they did not consider the variation of speed of light and gravitational constant in their model.

While on the surface it appears that fitting observation with a minimum number of parameters is more satisfactory than with a larger number of parameters, the models that can determine larger number of parameters from the same observation is more desirable in order go deeper in understanding the universe. A significant amount of work is required to understand how the new approach of this paper could determine other parameters of the universe and if it can resolve Hubble constant tension.

One may notice that the Hubble constant $H_0$ values in Table 2, based on SNe 1a data fit, are 1% higher for the VcG and VcGΛ models than its value for the ΛCDM model. We have put the same values in Table 3 for the VcG and VcGΛ models as we have not yet determined the same using the CMB temperature anisotropy spectrum. Thus $H_0$ numbers in Table 3 are not comparable and should therefore not be seen as if they resolve the $H_0$ tension.

It is even more staggering that the spatial curvature of the universe is strongly negative rather than flat and thus contradicts all the theories and observations derived assuming physical constants not varying. When analysing any observational data we need to be careful that it is not biased in favour of the physical constants that are constrained to their currently observed values, especially in view of the new definition of length dependent on the constant value of the speed of light as 299,792,458 meters per second (NIST 2020). As discussed below, physical constants other than $c$, $G$ and $\Lambda$ that are not directly involved in the models here, such as Planck constant $h$ and Boltzmann constant $k_B$ also vary. This makes it even more difficult to reach conclusions based on observations that directly or indirectly involve many physical constants.

It should also be pointed out that the matter density $\Omega_{m,0} = 0.528$, obtained by VcGΛ model is significantly higher than that estimated by the Planck mission $\Omega_{m,0} = 0.3153$ assuming ΛCDM cosmology (Planck Collaboration 2019). The Planck mission also estimated the baryon density $\Omega_{b,0} = 0.049$ and dark-matter density $\Omega_{dm,o} = 0.2607$ (Planck papers label it as $\Omega_c$). Since the baryon density has also been estimated directly by accounting the visible and non-visible baryonic matter, we can assume it to be valid for any cosmological model. Then the dark-matter density for the VcGΛ model $\Omega_{dm,o} = 0.479$ is 84% higher than the ΛCDM model. Cosmologist now will have more freedom to play with the dark-matter.

The results presented here may be considered a continuation of our previous work (Gupta 2019) wherein we explained three astrometric anomalies and the null results on the variation of $G$ and the fine structure constant using the quasi-phenomenological version of the current VcGΛ model. Since our current model reduces to the quasi-phenomenological model in the limit of the scale factor $a$ close to its current value 1, and since all these problems relate to the space where scale factor $a$ is close to 1, we conclude that the current model is also able to resolve these problems. Thus the findings of our previous work that the Planck constant $h$ varies just like the speed of light $c$ (Gupta 2019), inferred from the null results on the variation of the fine-structure constant $\alpha$, can also be considered valid under the VcGΛ model.

We have assumed in our work here that $k_B$ varies as $c^{1.25}$. This assumption leads to the redshift $z'(= z/e)$ for the surface of last scattering to be 400. This is confirmed from our results for the first peak of the CMB power spectrum and BAO acoustic scale which are within 3% of the standard ΛCDM model and observations. One could be sure that the results would only improve with the refinement of the model.

The variation of the Planck constant $\hbar$ has been the subject of several studies (e.g. Davies et al. 2002; Kentosh & Mohageg 2012; Mangano et al. 2015; de Gosson 2017; Massood-ul-Alam 2018). The interest in the variation of $\hbar$ has been mainly due to its occurrence in the fine structure constant $\alpha = \frac{1}{4\pi\varepsilon_0}\frac{e^2}{\hbar c}$. If $\alpha$ varies, then one could ask which constant occurring in its expression does vary. The variation of the Boltzmann constant has not enjoyed such attention. However, the variability of any constant can be associated with a dynamical field that evolves due to the action of a background potential which drives the field towards its minimum. Thus it is implicitly assumed that the field reached a stable minimum very early in the Big-Bang era resulting in the constancy of the physical constant associated with the field. An evolutionary background potential could therefore cause the field minimum to also evolve. This may also explain why all the physical constants might evolve and why studying any constant in isolation might not be prudent.

It would be interesting to see how some of the well-known constants vary when the fundamental constants are considered evolutionary. Consider for example the Rydberg constant $R_\infty = \frac{m_e e^4}{8\varepsilon_0^2 h^3 c}$ where $m_e$ is the rest mass of the electron, $e$ is the elementary charge, and $\varepsilon_0$ is the permittivity of space. If we consider $m_e$ and $e$, which belong to baryonic matter, to not vary (at least in comparison to the variation of the fundamental constants considered here), and since $\varepsilon_0 \propto 1/c^2$ and $h$ varies as $c$, we infer that $R_\infty$ does not vary. Another example can be taken as the Stefan-Boltzmann constant $\sigma = 2\pi^5 k_B^4 / 15 h^3 c^2$. Again, since $h$ varies as $c$ and $k_B$ varies as $c^{1.25}$, $\sigma$ does not vary. Similarly, the fine-structure constant, given by $\alpha = \frac{1}{4\pi\varepsilon_0}\frac{e^2}{\hbar c}$ will be immune to the variation of $\hbar$ and $c$. As already mentioned, it is the constancy of the fine-structure constant that led us to conclude in our previous work (Gupta 2019) that the Planck constant varies as the speed of light. Any variations in these constant would mean either our assumptions are wrong or $e$ and $m_e$ do vary to the extent of the measured variations in the Rydberg constant and the fine structure constant.

We could think of physical constants belonging to two different categories: constants that are independent of the Hubble expansion of the universe - e.g. the fine structure constant $\alpha$ and proton to electron mass ratio $\mu$; and the constants that may be tied to the Hubble expansion - e.g. the Newton's gravitational constant $G$, and possibly also the speed of light $c$, the Planck's constant $h$ and the Boltzmann constant $k_B$. The constants in the first category might be varying much more slowly, if varying at all, than those in the second category. The problem we see is that, in the absence of any knowledge about how different constants vary, one tends to ignore the possible variation of other constants in studying the variation of one constant. We have tried to study some constants in the second category and found that in many expressions and formulae they vary in such a way that they negate the variations of others and thus make their variations unobservable.

It should be explicitly stated that both the models VcG and VcGΛ are compliant with the work of Costa et al. (2019).

## 5. CONCLUSIONS

We have shown that the VcGΛ model, based on the covariant relativistic approach for including variation of fundamental physical constants can explain the cosmological observation considered in this work, as well or better than the standard ΛCDM model, with very few parameters. Most difficult to accept findings of the current work are (a) the evolutionary dark-energy density being negative, and (b) the universe having negative spatial curvature. Nonetheless, the VcGΛ model (i) fits the supernovae Ia observational data marginally better than the ΛCDM model; (ii) determines the first peak in the power spectrum of the cosmic microwave background temperature anisotropies at the multipole value of $l = 217.3$ against the observed value of 220 derived from ΛCDM model; (iii) calculates the age of the universe as 14.1 Gyr against the accepted value of 13.7 Gyr; and (iv) finds the BAO acoustic scale to be 145.2 Mpc against the observed value of 149.3. Also, the analysis presented here predicts the energy density peak in the blackbody frequency spectrum at $\nu_{max}$ to shift with time as $\frac{\dot{\nu}_{max}}{\nu_{max}} \approx 9 \times 10^{-11} \text{yr}^{-1}$. We therefore conclude that the VcGΛ model deserves to be considered seriously for further work. It will require collaboration to study more difficult cosmological problems such as fitting the power spectrum of CMB temperature anisotropies. Current codes such as CAMB (Lewis, Challinor & Lasenby 2000), CLASS (Lesgourgues 2011) and CMBAns (Das & Phan 20190 are unable to handle background and other changes required to properly handle VPC models. It is challenging undertaking to modify them asexpressed by the lead authors of these codes in private communications.

## ACKNOWLEDGEMENTS

This work has been supported by a generous grant from Macronix Research Corporation. Thanks are due to Madhav Singhal at the University of Western Ontario who used his python code to verify the SNe 1a data fit by Matlab curve fitting app. The author is grateful to Krishnakanta Bhattacharya at the Indian Institute of Technology, Guwahati, for his critical comments and helpful discussions. He remains thankful to Dan Scolnic at Duke University for providing the SNe 1a Pantheon Sample data used in this work, and to Santanu Das at the University of Wisconsin, Madison, for his comments and discussion on the use and modification of CMB codes. The author is grateful to Antony Lewis at the University of Sussex and Julien Lesgougues at Aachen University for advising how it might be possible to modify their CMB codes to include variable physical constants and non-flat universes. He wishes to express his gratitude to the reviewer of the paper whose very constructive critical comments helped improve the paper significantly.

## DATA AVAILABILITY

Data used is from references discussed in the manuscript (Scolnic, et al., 2018).